\documentclass[%
superscriptaddress,
showpacs,
amsmath,amssymb,
aps,
prc,
showkeys,
twocolumn,
]{revtex4-1}

\usepackage{amsmath,mathrsfs,amssymb}
\usepackage{eurosym}
\usepackage{graphicx,subfigure}
\usepackage{color}
\usepackage{indentfirst}
\usepackage{extarrows}
\usepackage{hyperref}
\hypersetup{colorlinks=true, citecolor=blue, urlcolor=blue, linkcolor=blue}

\begin{document}

\title{Effects of nonlocality of nuclear potentials on direct capture reactions}
\author{Yuan Tian}
\email[Correspondence author: ]{tyseraph@163.com}
\affiliation{China Institute of Atomic Energy, P.O. Box 275(10), Beijing 102413, China}

\author{D.Y. Pang}
\email[Correspondence author: ]{dypang@buaa.edu.cn}
\affiliation{School of Physics and Nuclear Energy Engineering, Beihang University, Beijing 100191, China}
\affiliation{Beijing Key Laboratory of Advanced Nuclear Materials and Physics, Beihang University, Beijing 100191, China}

\author{Zhong-yu Ma}
\affiliation{China Institute of Atomic Energy, P.O. Box 275(10), Beijing 102413, China}

\begin{abstract}

Calculations of the direct radiative capture reactions are made for the $^{48}$Ca$(n,\gamma)^{49}$Ca, $^7$Li$(n,\gamma)^8$Li and $^{12}$C$(p,\gamma)^{13}$N reactions with the Perey-Buck type nonlocal potentials using a potential model. Our results reproduce the experimental data reasonably well. From comparisons with results obtained by using local potentials, it is found that the cross sections of direct capture reactions may change by around 25\% due to the nonlcality of nuclear potentials.
\end{abstract}

\maketitle

\section{Introduction}

Radiative capture of nucleons at energies of astrophysical interest is one of the most important processes for nucleosynthesis. The nucleon capture can occur through both the compound nucleus formation and direct capture processes~\cite{Meyer-ARAA-1994, Goriely-PRL-2013, Banerjee-PRL-2013}. At low incident energies, the cross sections of the compound processes are usually very small because only a few excited states of the compound nuclei are involved. In these cases, the direct capture mechanism may be dominant. The direct radiative capture reactions, especially at the low energy region, play a crucial role in studies of big bang nucleosynthesis, main path stellar evolution, element synthesis at supernova sites, X-ray bursts etc., since these cross sections are often necessary for investigating the astrophysical entities~\cite{Rolfs-book,Burles-PRL-1999}. These necessitate reliable theoretical models of direct radiative capture reactions from low to high energies.

The direct radiative capture process represents a transition of the projectile-target system from an initial continuum state to a final bound state via interaction with the electromagnetic field. The reaction selects those projectiles from the appropriate partial waves with orbital angular momentum that can jump into the final orbits by emission of $\gamma$ ray of multipolarity $L$. In order to calculate the direct capture cross sections, one needs to solve the many-body problems for the bound and continuum states of relevance. There are several levels of difficulties in attacking this problem. Theories, such as the microscopic cluster model~\cite{Descouvemont-NPA-1988} and the R-matrix method~\cite{Kremer-PRL-1988}, have been developed to overcome these difficulties. However, the simplest solution is the potential model~\cite{Christy-NP-1961, Parker-PR-1963, Rolfs-NPA-1973, Baye-AnnPhys-1985, Bertulani-CPC-2003}. This model represents the initial and final states of the reaction system with the continuum/scattering and bound state wave functions, which are solutions of the two-body Schr\"{o}dinger equation with a potential in the center-of-mass of the projectile and the target nuclei. The cross sections are sensitive to these potentials.

In principle, the potentials that are responsible to the scattering and reactions of a projectile (nucleon or nucleus) with a target nucleus are nonlocal. Sources of the nonlocality in these effective potentials arise predominantly from antisymmetrization \cite{Canton-PRL-2005} and channel couplings \cite{Mahaux-1991}. In a folding model, the nonlocal character of the nucleon-nucleus optical potential is solely determined by the off-shell structure of NN $t$-matrix \cite{Elster-PRC-1997, Gennari-PRC-2018, Burrows-PRC-2018}. The nonlocality of nucleon-nucleus potentials can be naturally dealt with in momentum space \cite{Elster-PRC-1990, Hlophe-PRC-2013, Hlophe-PRC-2014}. In coordinate space, nonlocal potentials are often given in tabular forms with microscopic models \cite{Hao-PRC-2015}, which are not convenient to be used in usual nuclear reaction calculations and to be compared with results of different systems and different works. Because of these reasons, the separable form of nonlocal potentials proposed by Perey and Buck (PB) \cite{Perey-NP-1962}, which is parameterized with a range of nonlocality, is most widely used in various nuclear reaction calculations \cite{Timofeyuk-PRC-2013, Timofeyuk-PRL-2013, Titus-PRC-2014, Johnson-PRC-2014, Ross-PRC-2015, Titus-PRC-2016} although it may not represent the real structure of nonlocal potentials given by microscopic theories sufficiently well. Some systematic nonlocal nucleon-nucleus potentials have also been proposed with the Perey-Buck form \cite{Perey-NP-1962, Tian-2015-IJMPE}.

We adopt the Perey-Buck type nonlocal potential to study the effects of potential nonlocality to direct capture reactions. In Sec. II we briefly introduce the numerical method to solve the Schr\"{o}dinger equations with nonlocal potentials. We investigate the nonlocality effects on neutron and proton direct capture reactions in Sec. III. The cases reported here are the $^{48}$Ca$(n,\gamma)^{49}$Ca, $^7$Li$(n,\gamma)^8$Li and $^{12}$C$(p,\gamma)^{13}$N, reactions. A summary of the present work is presented in Sec. IV. 

\section{The Potential Model For Direct Radiative Capture Reactions}

With the potential model~\cite{Christy-NP-1961, Parker-PR-1963, Rolfs-NPA-1973, Baye-AnnPhys-1985, Bertulani-CPC-2003}, the cross sections for direct radiative capture reaction, $x(n,\gamma)a$ is: 
\begin{equation}\label{a1}
\begin{split}
    \sigma_{\pi L,J_b}^\textrm{d.c.}&=\frac{(2\pi)^3}{k^2}\left(\frac{E_{nx}+E_b}{\hbar c}\right)\frac{2(2I_a+1)}{(2I_n+1)(2I_x+1)}\\
    &\times\frac{L+1}{L[(2L+1)!!]^2}\sum_{J_cj_cl_c}(2J_c+1)\\
    &\times\left\{\begin{array}{ccc}
    j_c&J_c&I_x\\ J_b&j_b&L\end{array}\right\}^2|\langle l_bj_b||\mathscr{O}||l_cj_c\rangle|^2,
\end{split}
\end{equation}
where $\pi=E$ or $M$, which stands for electronic or magnetic transitions, respectively, $L$ is the multipoliarity of the emitting $\gamma$-ray. $\pmb{I}_n$, $\pmb{I}_x$ and $\pmb{I}_a$ are intrinsic spins of the cluster $n$ ($n$ can also be a nucleon), the nucleus $x$, and the composite nucleus $a=n+x$, respectively.
In the $n$-$x$ system, $\pmb{l}_c$ is the orbital angular momentum of $n$ in continuum states, and $\pmb{j}_c=\pmb{l}_c+\pmb{I}_n$. Similarly, $\pmb{l}_b$ is the orbital angular momentum of $n$ in bound states, and $\pmb{j}_b=\pmb{l}_b+\pmb{I}_n$. $\pmb{J}_c=\pmb{j}_c+\pmb{I}_x$ and $\pmb{J}_b=\pmb{j}_b+\pmb{I}_x$ are channel spins of the incident- and exit-channels (corresponding to continuum and bound states of the $n$-$x$ system). $E_b$ is the binding energy of $n$ in the bound states of $a$. $E_{nx}$ is the incident energy in the center-of-mass system and $\langle l_bj_b||\mathscr{O}||l_cj_c\rangle$ is the reduced matrix element, which can be expressed as a product of two factors:
\begin{equation}\label{a2}
    \langle l_bj_b||\mathscr{O}||l_cj_c\rangle=\tau_{b,c}A_{b,c},
\end{equation}
where
\begin{equation}\label{tau}
    \tau_{b,c}=\int u_b(r) r^L u_c(r)dr
\end{equation}
is the overlap integral of the radial parts of the scattering wave function, $u_c$, and the bound state wave function, $u_b$, of $n$ in the $n$+$x$ system. The factor $A_{b,c}$ denotes an angular momentum coupling coefficient~\cite{Rolfs-NPA-1973}.

The total direct capture cross section is obtained by adding all multipolarities and final spins of the bound state.
\begin{equation}\label{}
    \sigma^\textrm{d.c.}(E_{nx})=\sum_{LJ_b}(SF)_{J_b}\sigma_{LJ_b}^\textrm{d.c.}(E_{nx})
\end{equation}
where $(SF)_{J_b}$ are spectroscopic factors of $n$ in each of the bound states of $a$. For charged particles it is more convenient to use the astrophysical $S$-factors instead of the cross sections:
\begin{equation}\label{eq-Sfactor}
      S(E_{nx})=E_{nx}\sigma^\textrm{d.c.}(E_{nx})\exp[2\pi\eta(E_{nx})],
\end{equation}
where $\eta(E_{nx})$ is the Sommerfeld parameter:
\begin{equation}
	  \eta(E_{nx})=\frac{Z_nZ_xe^2}{\hbar}\left(\frac{\mu}{2E_{nx}}\right)^{1/2},
\end{equation}
$\mu$ is the reduced mass of $n$ in the $n$-$x$ system, and $Z_n$ and $Z_x$ are charge numbers of $n$ and $x$, respectively.

The radial wave functions $u_b$ and $u_c$ in Eq.~(\ref{tau}) are solutions of Schr\"{o}dinger equations. With local potentials, the radial part of the schr\"{o}dinger equation reads:
\begin{equation}\label{eq-schro-lc}
\begin{split}
    \frac{\hbar^2}{2\mu}&\left[\frac{d^2}{dr^2}-\frac{l(l+1)}{r^2}\right]u_{jl}(r)
    +[E-V_L(r)\\
    &-V_C(r)-(\pmb{I_n}\cdot\pmb{l})V_\textrm{so}(r)]u_{jl}(r)=0,\\
\end{split}
\end{equation}
where the energy $E$ is $E_b$ for a bound state, and is $E_{nx}$ for a scattering state, $\pmb{l}$ is the angular momentum of $n$ in the $n$-$x$ system, $\pmb{j}=\pmb{l}+\pmb{I_n}$, $V_L(r)$ and $V_\textrm{so}(r)$ are the central and the spin-orbital parts of the local potentials, respectively, and $V_\textrm{C}(r)$ is the Coulomb potential assuming a uniform charge distribution with a radius $R_\textrm{C}$:
\begin{equation}
\begin{array}{lll}
V_\textrm{C}(r)&=\displaystyle\frac{Z_nZ_xe^2}{2R_\textrm{C}}\left(3-\frac{r^2}{R_\textrm{C}^2}\right),&\hbox{for}~r\leq R_\textrm{C}\\
&=\displaystyle\frac{Z_nZ_xe^2}{r}. &\hbox{for}~r>R_\textrm{C}
\end{array}
\end{equation}
A Woods-Saxon and a derivative of Woods-Saxon form factors are used for $V_{L}(r)$ and $V_\textrm{so}(r)$, respectively, namely \cite{Bertulani-CPC-2003},
\begin{equation}\label{eq-Vl}
V_{L}(r)=V_{L}f_0(r),
\end{equation}
and 
\begin{equation}\label{eq-Vso}
V_\textrm{so}(r)=2V_\textrm{so}\frac{1}{r}\frac{d}{d r}f_\textrm{so}(r),
\end{equation}
where 
$$
f_i(r)=\left[1+\exp\left(\frac{r-R_i}{a_i}\right)\right]^{-1},
$$
$i=0$ and so labeling the central and spin-orbital terms, respectively, and $R_i=r_iA^{1/3}$ and $a_i$ are the radius and diffuseness parameters with $A$ being the atomic number of the target nucleus.

In this work, we only take the central part of the potential nonlocal and keep the spin-orbital and Coulomb terms local. Furthermore, the Perey-Buck form of nonlocality is adopted, with which the schr\"{o}dinger equation reads:
\begin{equation}\label{eq-schro-nlc}
	\begin{split}
		\frac{\hbar^2}{2\mu}&\left[\frac{d^2}{dr^2}-\frac{l(l+1)}{r^2}\right]u_{jl}(r)+\\
    	&[E-V_C(r)-(\pmb{I_n}\cdot\pmb{l})V_\textrm{so}(r)]u_{jl}(r)-\\
		&\int_0^\infty g_l(r,r')u_{jl}(r')dr'=0,
	\end{split}
\end{equation}
where \cite{Perey-NP-1962}
\begin{equation}\label{gl}
    g_l(r,r')=\frac{1}{\sqrt{\pi}\beta}\exp\left[-\left(\frac{r^2+r'^2}{\beta^2}\right)\right]2i^lzj_l(-iz)W(p),
\end{equation}
and $W(p)=V_\textrm{NL}f_0(p)$, $p=\frac{r+r'}{2}$, and $z=\frac{2rr'}{\beta^2}$. Here $\beta$ is the range of non-locality and $j_l(-iz)$ is the spherical Bessel functions of the $l$-th order.

The schr\"{o}dinger equation with a nonlocal potential is solved with iterations. We firstly find the solution, $u_j^{(0)}$, of the Schr\"{o}dinger equation with an initial \textit{local} potential $V_{\textrm{init}}$:
\begin{equation}\label{}
\begin{split}
    \frac{\hbar^2}{2\mu}\left[\frac{d^2}{dr^2}-\frac{l(l+1)}{r^2}\right]u_j^{(0)}(r)+[E-(V_{\textrm{init}}(r)+\\
    V_C(r)+(\pmb{I_n}\cdot\pmb{l})V_\textrm{so}(r)]u_j^{(0)}(r)=0
\end{split}
\end{equation}
An iteration is then made until a converged result is obtained. For the $i$-th iteraction, we have:
\begin{equation}\label{}
\begin{split}
    &\frac{\hbar^2}{2\mu}\left[\frac{d^2}{dr^2}-\frac{l(l+1)}{r^2}\right]u_j^{(i)}(r)+[E-(V_\textrm{init}(r)\\
    &+V_C(r)+(\pmb{I_n}\cdot\pmb{l})V_\textrm{so}(r))]u_j^{(i)}(r)\\
    &=\int g_l(r,r')u_j^{(i-1)}(r')dr'-V_\textrm{init}(r)u_j^{(i-1)}(r).
\end{split}
\end{equation}

\section{Results of numerical calculations}

In this section, we present our results of numerical calculations for direct capture reactions, $^{48}$Ca$(n,\gamma)^{49}$Ca, $^7$Li$(n,\gamma)^8$Li and $^{12}$C$(p,\gamma)^{13}$N within the ranges of incident energies from 0.01 MeV to 0.4 MeV, from 0.01 MeV to 2 MeV, and from 0 MeV to 1.2 MeV, respectively, which are interesting for nuclear astrophysical studies. With these energy ranges, all these three reactions are $E1$ dominant \cite{Krausmann-PRC-1996, Beer-PRC-1996, Rolfs-NPA-1974, Nesaraja-PRC-2001, Burtebaev-PRC-2008}.
We firstly find the local and nonlocal potential parameters for bound and scattering states of these reactions in section \ref{sec-potentials}. The direct capture reaction cross sections are then calculated using these potentials and compared with experimental data in section \ref{sec-xsecs}.

\subsection{Local and nonlocal potential parameters for bound and scattering states}\label{sec-potentials}

Usual Woods-Saxon form factors are assumed for these potentials. Empirical values are used for the radius parameters and the range of nonlocality, namely, $r_0=1.25$ fm and $\beta=0.85$ fm. The other parameters are allowed to vary to simultaneously reproduce the binding energies and the $s$-wave scattering lengths of the $n$-$x$ systems. For the $^{12}$C$(p,\gamma)^{13}$N reaction, a potential supporting the resonance state at 0.422 MeV is also found. The same is done for local potentials.

For the $^{48}$Ca$(n,\gamma)^{49}$Ca reaction, we consider one neutron capture into the ground ($\tfrac{3}{2}^-$) and the first excited states ($\tfrac{1}{2}^-$, 2.023 MeV) of $^{49}$Ca. These states are assumed to consist of an inert $^{48}$Ca core and a neutron in the $2p_{3/2}$ and $2p_{1/2}$ orbitals, respectively. One set of potential parameters are found to simultaneously reproduce the neutron binding energies in both states of $^{49}$Ca and the $s$-wave scattering length, $\alpha_0=0.36\pm0.09$ fm \cite{Raman-PRC-1989}. 
For the $^7$Li$(n,\gamma)^8$Li reaction, the neutron is assumed to be captured into the ground state ($2^+$) and the first excited state ($1^+$, 0.981 MeV) of $^8$Li. These states are assumed to consist of an inert $^7$Li core and a neutron in the $1p_{3/2}$ and $1p_{1/2}$ orbitals, respectively. For this reaction we have to find two sets of potential parameters, one set to simultaneously reproduce the ground state binding energy and the spin-state scattering length $a_+=-3.63\pm0.05$ fm (groups ``local--gs'' and ``nonlocal--gs'' for local and nonlocal potentials, corresponding to the spin 2 of the ground state of $^8$Li) and 
another set to simultaneously reproduce the $E_b$ in the first excited state of $^8$Li and the spin-state scattering length $a_-=0.87\pm0.07$ fm (groups ``local--1ex'' and ``nonlocal--1ex'' for local and nonlocal potentials, corresponding to the spin 1 of the first excited state of $^8$Li) \cite{Koester-ZPA-1983}.
For the $^{12}$C$(p,\gamma)^{13}$N reaction, the proton is assumed to be captured in the ground state of $^{13}$N, which is considered to consist of a inert $^{12}$C core and a proton in the $1p_{1/2}$ orbital. There is a resonance state in the $p$-$^{12}$C system at 0.422 MeV with spin-parity of $\tfrac{1}{2}^+$. We found two sets of potential parameters for these two states, labeled as ``local--dir'' and ``local--res'' for local potentials and ``nonlocal-dir'' and ``nonlocal-res'' for nonlocal potentials, respectively.
Parameters of these potentials are given in Table. \ref{tab-parameters}. 

\begin{table}[htbp]
    \centering
    \caption{Potential parameters found for the $^{48}$Ca$(n,\gamma)^{49}$Ca, $^7$Li$(n,\gamma)^8$Li and $^{12}$C$(p,\gamma)^{13}$N reactions. See the text for the details. The units of $V_0$ and $V_\textrm{so}$ are MeV and MeV fm$^2$, respectively, and those of $r_0$, $a_0$, $r_\textrm{so}$ and $a_\textrm{so}$ are femtometer.}
    \begin{tabular}{ccccccccc}
        \hline\hline
        target & group  & $V_0$    & $r_0$    & $a_0$    & $V_\textrm{so}$   & $r_\textrm{so}$   & $a_\textrm{so}$   & $\beta$ \\\hline
        $^{48}$Ca  
        &    local & 47.24 & 1.25  & 0.65  & 16.52 & 1.25  & 0.65  &  \\
        & nonlocal & 60.58 & 1.25  & 0.65  & 18.25 & 1.25  & 0.65  & 0.85 \\        
        $^7$Li   & local--gs & 40.35 & 1.25  & 0.773 & 10    & 1.25  & 0.65  &  \\
        & local--1ex & 46.52 & 1.25  & 0.603 & 10    & 1.25  & 0.65  &  \\
        & nonlocal--gs & 45.78 & 1.25  & 0.773 & 10    & 1.25  & 0.65  & 0.85 \\
        & nonlocal--1ex & 54.5  & 1.25  & 0.603 & 10    & 1.25  & 0.65  & 0.85 \\
        $^{12}$C   & local--dir & 41.66 & 1.25  & 0.65  & 10    & 1.25  & 0.65  &  \\
        & local--res & 54.44 & 1.25  & 0.65  & 10    & 1.25  & 0.65  &  \\
        & nonlocal--dir & 48.35 & 1.25  & 0.65  & 10    & 1.25  & 0.65  & 0.85 \\
        & nonlocal--res & 70.43 & 1.25  & 0.65  & 10    & 1.25  & 0.65  & 0.85 \\
        \hline\hline
    \end{tabular}%
    \label{tab-parameters}%
\end{table}%

\begin{figure*}
    \includegraphics[width=\textwidth]{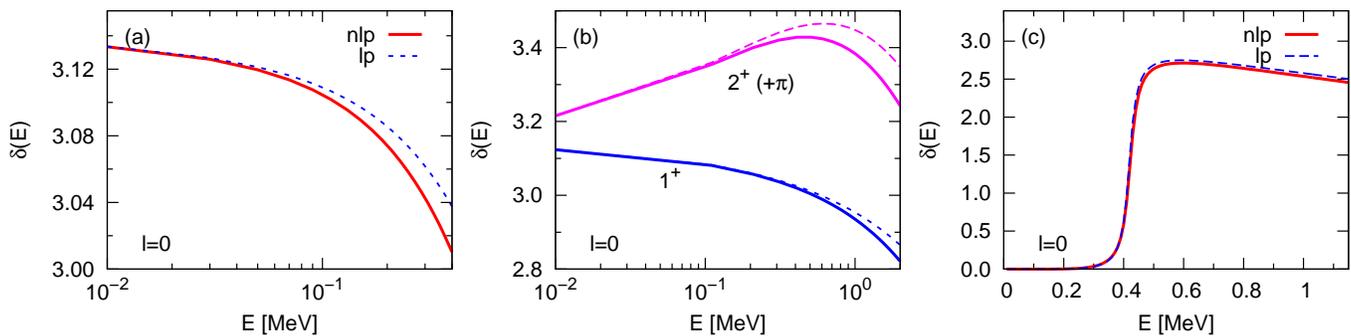}
    \caption{(Color online) $S$-wave phase shifts as functions of incident energies for neutron scattering from $^{48}$Ca (a), from $^{7}$Li (b), and for proton scattering from $^{12}$C (c) with the potentials listed in Table. \ref{tab-parameters}. Phase shifts obtained with nonlocal and local potentials are presented with solid and dashed curve, respectively. The $s$-wave phase shifts for the total spin $2^+$ of the $n$+$^7$Li system is shifted by $\pi$ for better visualization.}
    \label{fig-psh}
\end{figure*}

The phase shifts for neutron scattering from $^{48}$Ca and $^7$Li and for proton scattering from $^{12}$C with these potentials are plotted in Fig. \ref{fig-psh} as functions of incident energies. For simplicity, only phase shifts of the $s$-waves are presented, which are the most important partial waves for these reactions at energies below 1 MeV. Clearly, we see that these nonlocal potentials and their associated local counterparts are not phase equivalent. The differences in their phase shifts increase when the incident energy increases. Given the fact that, for each nucleon-target system, both local and nonlocal potentials are obtained by fitting the same binding energies and $s$-save scattering lengths, these results suggest that other properties of the nucleon-target systems, such as their effective ranges \cite{JiChen-PRC-2014, Beane-NPA-1998}, are needed to confine these potential parameters. We, however, do not endeavor to pursue exact phase equivalence of these local and nonlocal potentials in this work. With these potentials, we examine the effects of nuclear potential nonlocality to direct capture reactions in the following text.

\subsection{Direct capture reaction cross sections with local and nonlocal potentials}\label{sec-xsecs}

\begin{figure*}
    \includegraphics[width=\textwidth]{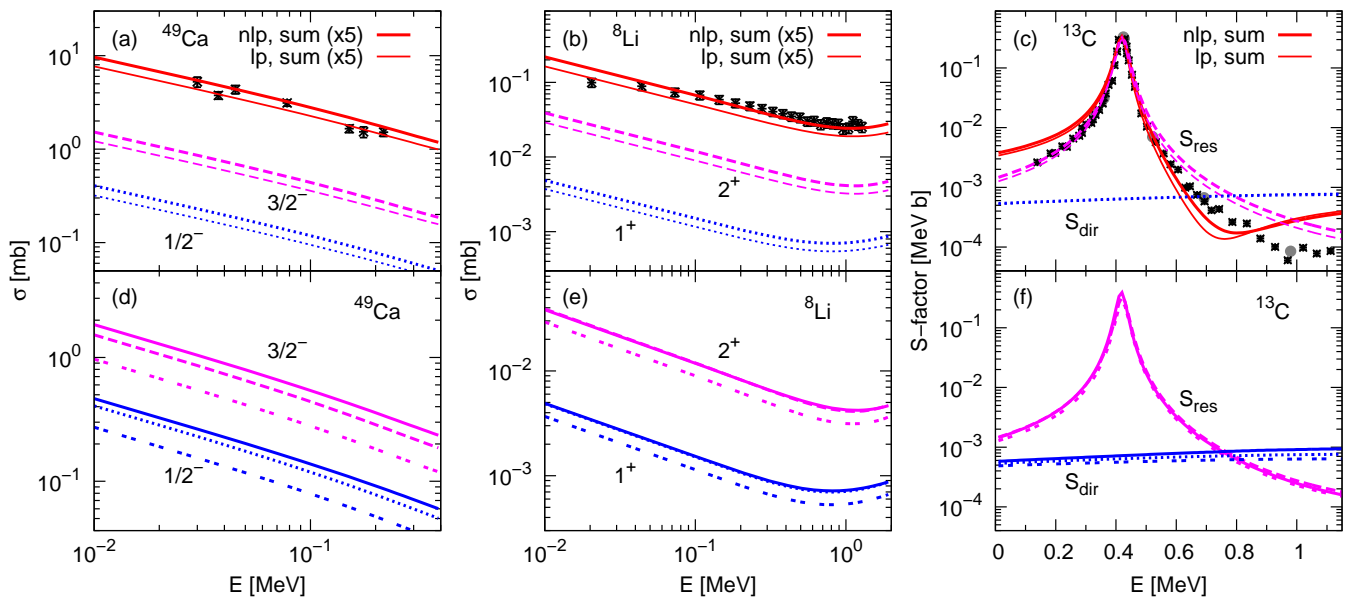}
    \caption{(Color online) Upper panels: comparisons between  experimental data and results of theoretical calculations using local and nonlocal potentials for reactions $^{48}$Ca(n,$\gamma$)$^{49}$Ca (a), $^{7}$Li($n,\gamma$)$^{8}$Li (b), and $^{12}$C($p,\gamma$)$^{13}$N (c). Contributions of transition to the ground (dashed curves) and the first excited sates (dotted curves) are shown explicitly for the $^{48}$Ca(n,$\gamma$)$^{49}$Ca and $^{7}$Li($n,\gamma$)$^{8}$Li reactions and their summed cross sections (multiplied by a factor of 5) are represented by solid curves. Results obtained with local and nonlocal potentials are discriminated by thin and thick curves, with labels ``nlp'' and ``lp'', respectively. Similarly, contributions of direct and resonant terms and their coherent sums are plotted for the $^{12}$C($p,\gamma$)$^{13}$N reaction. The symbols represents the experimental data for the $^{48}$Ca(n,$\gamma$)$^{49}$Ca reaction from Ref. \cite{Beer-PRC-1996}, for the $^{7}$Li($n,\gamma$)$^{8}$Li reaction from Ref. \cite{Izsak-PRC-2013} and for the $^{12}$C($p,\gamma$)$^{13}$N reaction from Refs. \cite{Rolfs-NPA-1974} (asterisk) and \cite{Burtebaev-PRC-2008} (solid circles). 
        Lower panels: comparisons of cross sections with potential nonlocality treated in both bound and scattering states (the same as in the upper panels) with those treated in the bound states only (solid curves) and in the scattering states only (double-dotted curves) for reactions $^{48}$Ca(n,$\gamma$)$^{49}$Ca (d), $^{7}$Li($n,\gamma$)$^{8}$Li (e), and $^{12}$C($p,\gamma$)$^{13}$N (f). See the text for details.}
    \label{fig-xsec}
\end{figure*}

Cross sections of the $^{48}$Ca$(n,\gamma)^{49}$Ca, $^7$Li$(n,\gamma)^8$Li and $^{12}$C$(p,\gamma)^{13}$N reactions are calculated using the aforementioned nonlocal potentials within the ranges of incident energies from 0.01 MeV to 0.4 MeV, from 0.01 MeV to 2 MeV, and from 0 MeV to 1.2 MeV, respectively. Their comparisons with the experimental data and with the results calculated using local potentials are also made. The results are shown in Fig. \ref{fig-xsec}. All calculations are made with a modified version of the computer code RADCAP \cite{Bertulani-CPC-2003}.

For the $^{48}$Ca$(n,\gamma)^{49}$Ca and $^7$Li$(n,\gamma)^8$Li reactions, processes of neutron capture to the ground and the first excited states of the composite nuclei are calculated separately. The summed cross sections are then obtained by adding the cross sections of these processes multiplied with their corresponding neutron spectroscopic factors (SFs). The same is done for both local and nonlocal potentials. 
The neutron SFs in the ground and the first excited states of
the composite nuclei are 0.72 and 0.86, respectively, for 
$^{49}$Ca \cite{Beer-PRC-1996} and are 0.87 and 0.48, respectively, for $^{8}$Li \cite{Nagai-PRC-2005}. 
From Fig. \ref{fig-xsec}, one sees that these summed cross sections reproduce the experimental data reasonably well. 
The cross sections with the nonlocal potentials are about 20\% larger than those with the local ones for the $^{48}$Ca$(n,\gamma)^{49}$Ca reaction. However, due to their large uncertainties, the experimental data of this reaction can not discriminate which theoretical result agrees better with them. The experimental data of the $^7$Li$(n,\gamma)^8$Li reaction, on the other hand, is shown to be better reproduced with nonlocal potentials, which is about 25\% larger than the cross sections with local potentials when the incident energy is below around 1 MeV.

For the $^{12}$C$(p,\gamma)^{13}$N reaction, the astrophysical $S$-factor defined in Eq. (\ref{eq-Sfactor}) is used instead of the cross sections. Description of the experimental data requires a coherent sum of the direct and resonance terms \cite{Rolfs-NPA-1974}:
\begin{equation}
\begin{split}
S(E)&=S_\textrm{dir}(E)+S_\textrm{res}(E)\\
&+2[S_\textrm{dir}(E)S_\textrm{res}(E)]^{1/2}\cos(\delta_\gamma),
\end{split}
\end{equation}
where $\delta_\gamma$ is the resonance phase shift given by
\begin{equation}
\delta_\gamma=\arctan\left[\frac{\Gamma(E)}{2(E-E_\gamma)}\right].
\end{equation}
The direct term, $S_\textrm{dir}$, is obtained when the bound and scattering state wave functions are calculated with the potential which was confined only with the ground state binding energy of proton in $^{13}$N (groups local--dir and nonlocal--dir in Table. \ref{tab-parameters}), while the resonance term, $S_\textrm{res}$, is obtained when the scattering wave functions are calculated with the potential that was adjusted to support the resonant state at 0.422 MeV (groups local--res and nonlocal--res in Table. \ref{tab-parameters}). 
The results are shown in Fig. \ref{fig-xsec} (c), from which, one sees that
the direct terms of the S-factor are almost identical with local and nonlocal potentials. The changes induced by the nonlocality of nuclear potentials manifest themselves in the resonance terms, $S_\textrm{res}$. For incident energies at the vicinity of the resonance energy and below, $S_\textrm{res}$ calculated with local and nonlocal potentials are very close to each other. Their difference increases with the increase of the incident energy. At around 1 MeV, they differ by around 25\%. The interference between $S_\textrm{dir}$ and $S_\textrm{res}$ are apparent and is seen to be important for the description of the experimental data, which is familiar as in Ref. \cite{Rolfs-NPA-1974}. At the vicinity of the resonance energy, the experimental data are rather well reproduced by the coherent sums of $S_\textrm{dir}$ and $S_\textrm{res}$ calculated with both nonlocal and local potentials. 
At above the resonant energies, however, they are underestimated below 0.85 MeV and are overestimated above 0.85 MeV by calculations with both potentials. The nonlocal potential improved the description to the experimental data between 0.55 and 0.8 MeV, but the calculated $S$-factor are still smaller than the experimental ones by around 50\%. Calculations with both local and nonlocal potentials also overestimated the experimental data at below the resonant energies.
In all these calculations, $S_\textrm{dir}$ is calculated with a proton spectroscopic factor of 0.81, which was determined with a $^{12}$C($^3$He,d)$^{13}$N reaction \cite{Sercely-NPA-1979}. The proton SF in the 0.422 MeV resonant state was then determined by matching the calculated $S(E)$ with the experimental $S$-factor at the resonant energy. The resulting SF of the resonant state is 0.36. It is very close to the value (SF=0.35) obtained in Ref. \cite{Huang-ADNDT-2010}.

The nonlocality of nuclear potentials affects both the bound and the scattering wave functions as compared with the wave functions calculated using local potentials although both potentials reproduce the same binding energies and $s$-wave scattering lengths. It is interesting to see the effects to the direct capture reactions from changes in the bound and scattering wave functions separately. In the bottom panels of Fig.~\ref{fig-xsec} we compare results of calculations with potential nonlocality treated in both bound and scattering wave functions (dashed and dotted curves) with those calculated with nonlocality treated only in the bound states (solid curves) or only in the scattering states (double-dotted curves) for the three reactions. These results show that the effects of potential nonlocality in bound and scattering state wave functions affect the direct capture reactions differently and they interfere constructively in the $^7$Li$(n,\gamma)^8$Li
reaction and destructively in the  $^{48}$Ca$(n,\gamma)^{49}$Ca and $^{12}$C$(p,\gamma)^{13}$N reactions.

\section{summary}

Effects of potential nonlocality in direct radiative capture reactions are studied with $^{48}$Ca$(n,\gamma)^{49}$Ca, $^7$Li$(n,\gamma)^8$Li and $^{12}$C$(p,\gamma)^{13}$N reaction at low energies with a potential model. Parameters of Perey-Buck type nonlocal potentials are found for these reaction systems and theoretical cross sections with these potentials are compared with experimental data. Our results show that the reproduction to the experimental data is improved when nonlocal potentials are used for the $^7$Li$(n,\gamma)^8$Li and $^{12}$C$(p,\gamma)^{13}$N reactions. 
A change of cross sections up to around 25\% is found for these three reaction. 
The effects of potential nonlocality in the bound and continuum state wave functions are found to affect the direct capture reactions differently and they interfere, which suggests that potential nonlocality should be treated simultaneously for both bound and continuum state in direct capture reactions.

\begin{acknowledgments}
This work is supported by the National Natural Science Foundation of China (Grants Nos. U1432247, 11775013, 11305270, 11465005, and U1630143) and the national key research and development program (2016YFA0400502). This work has been supported by the IAEA Coordinated Research Project F41032 (Grant No. 20466).
\end{acknowledgments}

\bibliographystyle{apsrev4-1}
\bibliography{ytian}

\end{document}